\documentclass[a4paper,10pt]{article}
\usepackage{amssymb}
\usepackage{amsmath}
\usepackage{epsfig}
\usepackage{subfigure}
\usepackage{graphics}
\usepackage{tikz}
\usepackage{latexsym}
\usepackage{rotating}
\usepackage{titlesec}
\newcommand{\squeezeup}{\vspace{-6mm}}
\usetikzlibrary{matrix}

\title{Soliton solutions of the shifted nonlocal NLS and MKdV equations}

\author{Metin G\"{u}rses \thanks{gurses@fen.bilkent.edu.tr}\\
{\small Department of Mathematics, Faculty of Science}\\
{\small Bilkent University, 06800 Ankara - Turkey}\\
Asl{\i} Pekcan \thanks{aslipekcan@hacettepe.edu.tr} \\
{\small Department of Mathematics, Faculty of Science} \\
{\small Hacettepe University, 06800 Ankara - Turkey}
}

\date{\nonumber}
\begin{document}
\maketitle
\date{\nonumber}

\begin{abstract}
We find one- and two-soliton solutions of shifted nonlocal NLS and MKdV equations. We discuss the singular structures of these soliton solutions and present some of the graphs of them.\\


\end{abstract}

\section{Introduction}

In the last decade we have witnessed some new nonlocal reductions of the system of equations and hence new nonlocal integrable equations \cite{AbMu1}-\cite{AbMu4}.
Soliton solutions of these equations; for instance, nonlocal nonlinear Schr\"{o}dinger equations (NLS) \cite{AbMu1}-\cite{zyan},
nonlocal modified Korteweg-de Vries (mKdV) equations \cite{AbMu2}-\cite{chen}, \cite{GurPek3}, \cite{GurPek2}-\cite{ma},  nonlocal sine-Gordon (SG) equations \cite{AbMu2}-\cite{chen}, \cite{aflm}, nonlocal Davey-Stewartson equations \cite{RZFH}-\cite{Rao}, and so on \cite{GursesFK}-\cite{hydro} have been discussed.

In particular in \cite{GurPek1}, \cite{GurPek3}, and \cite{GurPek2} we studied the coupled nonlinear Schr\"{o}dinger (NLS) system
\begin{align}
&aq_t=\frac{1}{2}q_{xx}-q^2r,\label{NLS1}\\
&ar_t=-\frac{1}{2}r_{xx}+r^2q,\label{NLS2}
\end{align}
and the coupled modified Korteweg-de Vries (MKdV) system
\begin{align}
&aq_t=-\frac{1}{4}q_{xxx}+\frac{3}{2}rqq_x,\label{MKdV1}\\
&ar_t=-\frac{1}{4}r_{xxx}+\frac{3}{2}rqr_x,\label{MKdV2}
\end{align}
where $a$ is any constant, and their nonlocal reductions. We used the Hirota bilinear method to obtain soliton solutions of those equations. In our method we first find soliton solutions of the system of equations then using the reduction formulas we obtain some relations on the solution parameters. We obtained one-soliton solutions of NLS and MKdV systems as \cite{GurPek1}, \cite{GurPek3}, \cite{GurPek2}
\begin{equation}\label{onesol}
\displaystyle q(x,t)=\frac{e^{k_1x+\omega_1t+\delta_1}}{1-\frac{1}{(k_1+k_2)^2}e^{(k_1+k_2)x+(\omega_1+\omega_2)t+\delta_1+\delta_2}}, \quad \quad r(x,t)=\frac{e^{k_2x+\omega_2t+\delta_2}}{1-\frac{1}{(k_1+k_2)^2}e^{(k_1+k_2)x+(\omega_1+\omega_2)t+\delta_1+\delta_2}},
\end{equation}
where the dispersion relations $ \omega_1=\frac{k_1^2}{2a}$, $ \omega_2=-\frac{k_2^2}{2a}$ hold for NLS system, and $ \omega_j=-\frac{k_j^3}{4a}$, $j=1, 2$ for MKdV system. Here $k_{1}$, $k_{2}$, $\delta_{1},$ and $\delta_{2}$ are arbitrary complex numbers. In addition to this, we also derived two-soliton solutions of NLS and MKdV systems as
\begin{align}
\displaystyle& q(x,t)=\frac{e^{\theta_1}+e^{\theta_2}+A_1e^{\theta_1+\theta_2+\eta_1}+A_2e^{\theta_1+\theta_2+\eta_2}}
{1+e^{\theta_1+\eta_1+\alpha_{11}}+e^{\theta_1+\eta_2+\alpha_{12}}+e^{\theta_2+\eta_1+\alpha_{21}}+e^{\theta_2+\eta_2+\alpha_{22}}
+Me^{\theta_1+\theta_2+\eta_1+\eta_2}},\label{twosolq(x,t)}\\
&r(x,t)=\frac{e^{\eta_1}+e^{\eta_2}+B_1e^{\theta_1+\eta_1+\eta_2}+B_2e^{\theta_2+\eta_1+\eta_2}}
{1+e^{\theta_1+\eta_1+\alpha_{11}}+e^{\theta_1+\eta_2+\alpha_{12}}+e^{\theta_2+\eta_1+\alpha_{21}}+e^{\theta_2+\eta_2+\alpha_{22}}
+Me^{\theta_1+\theta_2+\eta_1+\eta_2}},\label{twosolr(x,t)}
\end{align}
where $ \theta_i=k_ix+\omega_it+\delta_i$, $ \eta_i=l_ix+m_it+\alpha_i$, $i=1, 2$,
\begin{equation}
\displaystyle e^{\alpha_{ij}}=-\frac{1}{(k_i+l_j)^2},\, 1\leq i,j\leq 2,
\end{equation}
\begin{equation}\label{gamma_ibeta_i}
\displaystyle A_i=-\frac{(k_1-k_2)^2}{(k_1+l_i)^2(k_2+l_i)^2}, \quad B_i=-\frac{(l_1-l_2)^2}{(l_1+k_i)^2(l_2+k_i)^2},\quad i=1, 2,
\end{equation}
\begin{equation}\label{M}
\displaystyle M=\frac{(k_1-k_2)^2(l_1-l_2)^2}{(k_1+l_1)^2(k_1+l_2)^2(k_2+l_1)^2(k_2+l_2)^2},
\end{equation}
with the following dispersion relations
\begin{equation}\label{NLSdispersiontwo}
\omega_j=\frac{k_j^2}{2a},\quad  m_j=-\frac{l_j^2}{2a},\quad j=1, 2.
\end{equation}
and
\begin{equation}\label{MKdVdispersiontwo}\displaystyle
\omega_j=-\frac{k_j^3}{4a}, \quad m_j=-\frac{l_j^3}{4a}, \quad j=1, 2,
\end{equation}
hold for NLS system (\ref{NLS1}) and (\ref{NLS2}),
and MKdV system (\ref{MKdV1}) and (\ref{MKdV2}), respectively. Here $k_{i}$, $l_{i}, \delta_{i}$, and $\alpha_{i}$, $i=1, 2$ are arbitrary complex numbers.

By the nonlocal reductions; $r(x,t)=k q(\varepsilon_1x,\varepsilon_2t)$ and $r(x,t)=k \bar{q}(\varepsilon_1x,\varepsilon_2t)$, $\varepsilon_1^2=\varepsilon_2^2=1$, the coupled NLS system (\ref{NLS1}) and (\ref{NLS2}), and
the coupled MKdV system (\ref{MKdV1}) and (\ref{MKdV2}) reduce consistently to nonlocal NLS and nonlocal MKdV equations, respectively. Here bar notation stands for the complex conjugation. By using the above soliton solutions for $r(x,t)$ and $q(x,t)$ and the reduction equations we have obtained one- and two-soliton solutions of nonlocal NLS and nonlocal MKdV equations.

Recently, Ablowitz and Musslimani \cite{AbMu4} have found new nonlocal reductions (shifted nonlocal reductions); $r(x,t)=k \bar{q}(x_{0}-x,t)$, $r(x,t)=k q(x,-t+t_{0})$, $r(x,t)=k q(x_{0}-x,t_{0}-t)$, and $r(x,t)=k \bar{q}(x_{0}-x,t_{0}-t)$ where $x_{0}$ and $t_{0}$ are arbitrary real constants. By using these reductions the NLS and MKdV systems reduce to the shifted nonlocal NLS and MKdV equations, respectively. These shifted nonlocal equations reduce back to usual unshifted nonlocal forms of them when we take $x_0=t_0=0$. To find soliton solutions of these new integrable systems we follow the same approach. We use the above one- and two-soliton solutions of the systems of equations and using the shifted reduction formulas we find the soliton solutions of the shifted nonlocal NLS and MKdV equations.

In the following section we present all possible shifted nonlocal equations derivable from the NLS and MKdV systems. Then in the third section we give the soliton solutions of these shifted nonlocal equations.

\section{Shifted nonlocality}

Starting with the systems of equations (\ref{NLS1}), (\ref{NLS2}), and (\ref{MKdV1}), (\ref{MKdV2}) we find some new nonlocal integrable equations. The shifted nonlocal reductions have been given by Ablowitz and Musslimani \cite{AbMu4}.

\subsection{Shifted nonlocal NLS equations}

\noindent \textbf{i)}\, $r(x,t)=kq(x,-t+t_0)$, $k, t_0 \in \mathbb{R}$

Real reverse time shifted nonlocal NLS equation is
\begin{equation}\label{realtimeNLS}
aq_t(x,t)=\frac{1}{2}q_{xx}(x,t)-kq^2(x,t)q(x,-t+t_0).
\end{equation}

\noindent \textbf{ii)}\, $r(x,t)=kq(-x+x_0,-t+t_0)$, $k, x_0, t_0 \in \mathbb{R}$

Real reverse space-time shifted nonlocal NLS equation is
\begin{equation}\label{realspacetimeNLS}
aq_t(x,t)=\frac{1}{2}q_{xx}(x,t)-kq^2(x,t)q(-x+x_0,-t+t_0).
\end{equation}

\noindent \textbf{iii)}\, $r(x,t)=k\bar{q}(-x+x_0,t)$, $k, x_0 \in \mathbb{R}$

Complex reverse space shifted nonlocal NLS equation is
\begin{equation}\label{complexspaceNLS}
aq_t(x,t)=\frac{1}{2}q_{xx}(x,t)-kq^2(x,t)\bar{q}(-x+x_0,t),
\end{equation}
where $a=-\bar{a}$.

\noindent \textbf{iv)}\, $r(x,t)=k\bar{q}(x,-t+t_0)$, $k, t_0 \in \mathbb{R}$

Complex reverse time shifted nonlocal NLS equation is
\begin{equation}\label{complextimeNLS}
aq_t(x,t)=\frac{1}{2}q_{xx}(x,t)-kq^2(x,t)\bar{q}(x,-t+t_0),
\end{equation}
where $a=\bar{a}$.

\noindent \textbf{v)}\, $r(x,t)=k\bar{q}(-x+x_0,-t+t_0)$, $k, x_0, t_0 \in \mathbb{R}$

Complex reverse space-time shifted nonlocal NLS equation is
\begin{equation}\label{complexspacetimeNLS}
aq_t(x,t)=\frac{1}{2}q_{xx}(x,t)-kq^2(x,t)\bar{q}(-x+x_0,-t+t_0),
\end{equation}
where $a=\bar{a}$.

\subsection{Shifted nonlocal MKdV equations}

\noindent \textbf{i)}\, $r(x,t)=kq(-x+x_0,-t+t_0)$, $k, x_0, t_0 \in \mathbb{R}$

Real reverse space-time shifted nonlocal MKdV equation is
\begin{equation}\label{realspacetimeMKdV}
aq_t(x,t)=-\frac{1}{4}q_{xxx}(x,t)+\frac{3}{2}kq(x,t)q_x(x,t)q(-x+x_0,-t+t_0).
\end{equation}

\noindent \textbf{ii)}\, $r(x,t)=k\bar{q}(-x+x_0,t)$, $k, x_0 \in \mathbb{R}$

Complex reverse space shifted nonlocal MKdV equation is
\begin{equation}\label{complexspaceMKdV}
aq_t(x,t)=-\frac{1}{4}q_{xxx}(x,t)+\frac{3}{2}kq(x,t)q_x(x,t)\bar{q}(-x+x_0,t),
\end{equation}
where $a=-\bar{a}$.

\noindent \textbf{iii)}\, $r(x,t)=k\bar{q}(x,-t+t_0)$, $k, t_0 \in \mathbb{R}$

Complex reverse time shifted nonlocal MKdV equation is
\begin{equation}\label{complextimeMKdV}
aq_t(x,t)=-\frac{1}{4}q_{xxx}(x,t)+\frac{3}{2}kq(x,t)q_x(x,t)\bar{q}(x,-t+t_0),
\end{equation}
where $a=-\bar{a}$.

\noindent \textbf{iv)}\, $r(x,t)=k\bar{q}(-x+x_0,-t+t_0)$, $k, x_0, t_0 \in \mathbb{R}$

Complex reverse space-time shifted nonlocal MKdV equation is
\begin{equation}\label{complexspacetimeMKdV}
aq_t(x,t)=-\frac{1}{4}q_{xxx}(x,t)+\frac{3}{2}kq(x,t)q_x(x,t)\bar{q}(-x+x_0,-t+t_0),
\end{equation}
where $a=\bar{a}$.

\section{Soliton solutions of shifted nonlocal NLS and MKdV equations}

In \cite{GurPek1}, \cite{GurPek3}, \cite{GurPek2} we obtained one-, two-, and three-soliton solutions of the coupled NLS system (\ref{NLS1}) and (\ref{NLS2}) and MKdV system (\ref{MKdV1}) and (\ref{MKdV2}) by using the Hirota bilinear method. Since NLS and MKdV systems belong to the same hierarchy that is they both have the same recursion operator, the form of the soliton solutions obtained through the Hirota method is the same except the dispersion relations.

\subsection{One-soliton solutions}

By using Type 1 and Type 2 approaches with the reduction formulas we also obtain one-soliton solutions of the shifted nonlocal NLS and MKdV equations.
Note that while Type 1 is based on equating numerators and denominators of the functions $r(x,t)$ and $q(x,t)$ separately, Type 2 approach is based on the
cross multiplication \cite{GurPek1}, \cite{GurPek3}, \cite{GurPek2}.

Here we will also compare the one-soliton solutions of the unshifted nonlocal NLS and MKdV equations with the solutions of the shifted ones.

\noindent \textbf{i)}\, $r(x,t)=kq(x,-t+t_0)$, $k, t_0 \in \mathbb{R}$.\\

Under this reduction, if we use Type 1 we get the following constraints:
\begin{equation}
k_1=k_2,\quad e^{\delta_2}=ke^{\delta_1+\omega_1t_0},
\end{equation}
giving $\omega_2=-\omega_1$ for the coupled NLS system (\ref{NLS1}) and (\ref{NLS2}). Hence we get one-soliton solution of real reverse time shifted nonlocal NLS equation (\ref{realtimeNLS}) as
\begin{equation}\displaystyle
q(x,t)=\frac{e^{k_1x+\frac{k_1^2}{2a}t+\delta_1}}{1-\frac{k}{4k_1^2}e^{2k_1x+2\delta_1+\frac{k_1^2}{2a}t_0}}.
\end{equation}
For real parameters with $k<0$, the above solution is nonsingular. If we compare
the one-soliton solution of the real reverse time unshifted nonlocal NLS, say $u(x,t)$, with the above solution we get
$q(x,t)\neq u(x, t+\kappa(t_0))$ for any nonzero function $\kappa$. Note that Type 2 approach gives the same solution for (\ref{realtimeNLS}). Let $k_1=\alpha_1+i\beta_1$, $e^{\delta_1}=\alpha_2+i\beta_2$, and $a=\alpha_3+i\beta_3$, $\alpha_j, \beta_j \in \mathbb{R}$ for $j=1, 2, 3$. From the above solution we get
\begin{equation}\label{onesolCasei}\displaystyle
|q(x,t)|^2=\frac{2e^{\rho t-\xi_1t_0}(\alpha_2^2+\beta_2^2)}{k\sigma_1[\cosh(2\alpha_1x+\xi_1t_0+\delta)+\sigma_1\sin(2\beta_1x+\xi_2t_0-\omega_0)]},
\end{equation}
where
\begin{align}\displaystyle
&\xi_1=\frac{\alpha_1^2\alpha_3+2\alpha_1\beta_1\beta_3-\beta_1^2\alpha_3}{2(\alpha_3^2+\beta_3^2)},\,\, \xi_2=\frac{\beta_1^2\beta_3+2\alpha_1\beta_1\alpha_3-\alpha_1^2\beta_3}{2(\alpha_3^2+\beta_3^2)},\\
&\rho=\frac{\alpha_3(\alpha_1^2-\beta_1^2)+2\alpha_1\beta_1\beta_3}{\alpha_3^2+\beta_3^2},\, \, \omega_0=\arccos{\Big(\frac{2\alpha_2\beta_2(\alpha_1^2-\beta_1^2)
-2\alpha_1\beta_1(\alpha_2^2-\beta_2^2)}{(\alpha_1^2+\beta_1^2)(\alpha_2^2+\beta_2^2)}\Big)},\\
\end{align}
and $\delta=\ln\Big(\sigma_1k \frac{(\alpha_2^2+\beta_2^2)}{4(\alpha_1^2+\beta_1^2)}\Big)$. Here if $k<0$, $\sigma_1=-1$, and $k>0$, $\sigma_1=1$. The solution (\ref{onesolCasei}) is nonsingular if $-\frac{\beta_1(\xi_1t_0+\delta)}{\alpha_1}+\xi_2t_0-\omega_0\neq \frac{(4n+1)\pi}{2}$ when $k<0$, and
 $-\frac{\beta_1(\xi_1t_0+\delta)}{\alpha_1}+\xi_2t_0-\omega_0\neq \frac{(4n+3)\pi}{2}$ when $k>0$ for $n\in \mathbb{N}$.\\

\noindent \textbf{ii)}\, $r(x,t)=kq(-x+x_0,-t+t_0)$, $k, x_0, t_0 \in \mathbb{R}$.\\

When we use Type 1 approach in this case for both NLS and MdV systems we get $k_1=-k_2$ giving trivial solution $q(x,t)=0$. Therefore we use Type 2 and get the following constraints:
\begin{equation}
e^{\delta_1}=\sigma_1\frac{i(k_1+k_2)}{\sqrt{k}e^{\frac{k_1x_0+\omega_1t_0}{2}}},\quad e^{\delta_2}=\sigma_2\frac{i\sqrt{k}(k_1+k_2)}{e^{\frac{k_2x_0+\omega_2t_0}{2}}},
\end{equation}
where $\sigma_j=\pm 1$, $j=1, 2$. Therefore one-soliton solutions of real reverse space-time shifted nonlocal NLS (\ref{realspacetimeNLS}) and MKdV (\ref{realspacetimeMKdV}) equations are
\begin{equation}
q(x,t)=\frac{i\sigma_1e^{k_1x+\omega_1t}e^{\frac{-(k_1x_0+\omega_1t_0)}{2}}(k_1+k_2)}{[1+\sigma_1\sigma_2e^{(k_1+k_2)x+(\omega_1+\omega_2)t}
e^{-\frac{(k_1+k_2)}{2}x_0-\frac{(\omega_1+\omega_2)}{2}t_0}]\sqrt{k}},
\end{equation}
where $\omega_1=\frac{k_1^2}{2a}$, $\omega_2=-\frac{k_2^2}{2a}$ for NLS and $\omega_j=-\frac{k_j^3}{4a}$, $j=1, 2$ for MKdV equations. Here if $\sigma_1\sigma_2>0$ the solution is nonsingular. It is clear that one-soliton solutions of real reverse space-time unshifted nonlocal NLS and MKdV
equations, say $u(x,t)$, are related to the above solution by $q(x,t)=u(x-\frac{x_0}{2},t-\frac{t_0}{2})$.

\noindent \textbf{Example 1.} Consider the solution parameters as $k_1=i, k_2=1+\frac{i}{2},\sigma_1=\sigma_2=1, k=-1, a=\frac{1}{4}, x_0=-1, t_0=2$.
Then soliton solution of the real reverse space-time shifted nonlocal MKdV equation (\ref{realspacetimeMKdV}) is
\begin{equation}
|q(x,t)|^2=\frac{13e^{-\phi}}{8[\cosh(\phi)+\cos(\frac{3}{2}\phi)]},
\end{equation}
 where $\phi=x-\frac{1}{4}t+\frac{3}{4}$. The graph of this solution is given in Figure 1.
\begin{center}
\begin{figure}[h!]
\centering
\begin{minipage}[t]{1\linewidth}
\centering
\includegraphics[angle=0,scale=.30]{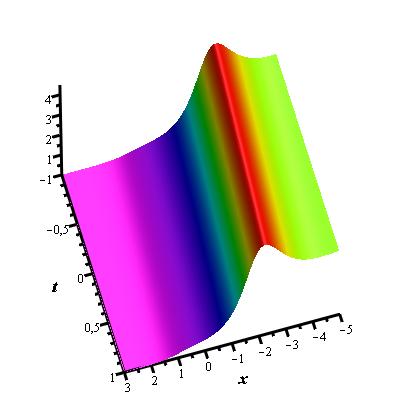}
\caption{A complexiton solution for (\ref{realspacetimeMKdV})
with the parameters $k_1=i, k_2=1+\frac{i}{2},\sigma_1=\sigma_2=1, k=-1, a=\frac{1}{4}, x_0=-1, t_0=2$.}
\end{minipage}%
\end{figure}
\end{center}

\squeezeup

\noindent \textbf{iii)}\, $r(x,t)=k\bar{q}(-x+x_0,t)$, $k, x_0 \in \mathbb{R}$.\\

For this case, when we use Type 1 we get
\begin{equation}
k_2=-\bar{k}_1,\quad e^{\delta_2}=ke^{\bar{k}_1x_0+\bar{\delta}_1}
\end{equation}
yielding $\omega_2=\bar{\omega}_1$. Hence we get one-soliton solutions of complex reverse space shifted nonlocal NLS (\ref{complexspaceNLS})
and MKdV (\ref{complexspaceMKdV}) as
\begin{equation}\displaystyle
q(x,t)=\frac{e^{k_1x+\omega_1t+\delta_1}}{1-\frac{k}{(k_1-\bar{k}_1)^2}e^{(k_1-\bar{k}_1)x+(\omega_1+\bar{\omega}_1)t
+\delta_1+\bar{\delta}_1+\bar{k}_1x_0}},
\end{equation}
where $\omega_1=\frac{k_1^2}{2a}$ for NLS and $\omega_1=-\frac{k_1^2}{4a}$ for MKdV equations. Here $a$ is a pure imaginary number. Let $k_1=\alpha_1+i\beta_1$, $\omega_1=\alpha_2+i\beta_2$, $e^{\delta_1}=\alpha_3+i\beta_3$, and $a=i\alpha_4$, $\alpha_j\in \mathbb{R}$ for $1\leq j\leq 4$, and
$\beta_j \in \mathbb{R}$ for $1\leq j\leq 3$. For NLS we have $\alpha_2=\frac{\alpha_1\beta_1}{\alpha_4}, \beta_2=\frac{(\beta_1^2-\alpha_1^2)}{2\alpha_4}$ and for MKdV $\alpha_2=\frac{(\beta_1^3-3\alpha_1^2\beta_1)}{4\alpha_4}, \beta_2=\frac{(\alpha_1^3-3\alpha_1\beta_1^2)}{4\alpha_4}$. If $\alpha_1\neq 0 $ the above solution of  complex reverse space shifted nonlocal NLS (\ref{complexspaceNLS})
and MKdV equation (\ref{complexspaceMKdV}) becomes
\begin{equation}\label{caseiiitype1ONE}
|q(x,t)|^2=\frac{2\beta_1^2e^{2\alpha_1x-\alpha_1x_0}}{k\sigma_1[\cosh(2\alpha_2t+\alpha_1x_0+\delta)+\sigma_1\cos(2\beta_1x-\beta_1x_0)]},
\end{equation}
where $\delta=\ln\Big(\frac{k\sigma_1(\alpha_3^2+\beta_3^2)}{4\beta_1^2}\Big)$. Here if $k<0$, $\sigma_1=-1$, and $k>0$, $\sigma_1=1$. If $k<0$, the above solution is singular at $(x,t)=(\frac{2n\pi+\beta_1x_0}{2\beta_1}, -\frac{(\alpha_1x_0+\delta)}{2\alpha_2})$, $n\in \mathbb{N}$. If $k>0$, then the solution is singular at $(x,t)=(\frac{(2n+1)\pi+\beta_1x_0}{2\beta_1}, -\frac{(\alpha_1x_0+\delta)}{2\alpha_2})$, $n\in \mathbb{N}$.

If $\alpha_1=0$ then $\alpha_2=0$ for NLS and the solution (\ref{caseiiitype1ONE}) of  complex reverse space shifted nonlocal NLS (\ref{complexspaceNLS}) becomes
\begin{equation}\label{iii-alpha_1=0-NLS}
|q(x,t)|^2=\frac{2\beta_1^2}{k\sigma_1[\cosh(\delta)+\sigma_1\cos(2\beta_1x-\beta_1x_0)]}.
\end{equation}
This solution is nonsingular if $\delta\neq 0$ i.e. $k\sigma_1(\alpha_3^2+\beta_3^2)\neq 4\beta_1^2$. For MKdV, in the case of $\alpha_1=0$ we have $\beta_2=0$ and the solution (\ref{caseiiitype1ONE}) of  complex reverse space shifted nonlocal MKdV (\ref{complexspaceMKdV}) becomes
\begin{equation}\label{iii-alpha_1=0-MKdV}
|q(x,t)|^2=\frac{2\beta_1^2}{k\sigma_1[\cosh(2\alpha_2t+\delta)+\sigma_1\cos(2\beta_1x-\beta_1x_0)]}.
\end{equation}
The above solution is nonsingular if $\alpha_2\delta>0$ for $t\geq 0$.\\

\noindent \textbf{Example 2.} Take the parameters of the solution (\ref{iii-alpha_1=0-NLS}) as $k=\sigma_1=\beta_1=\alpha_3=\beta_3=1$. It turns to be
a nonsingular periodic wave solution
\begin{equation}
|q(x,t)|^2=\frac{2}{\frac{5}{4}+\cos(2x-x_0)}.
\end{equation}
For $x_0=4$ we give the graph of the solution of complex reverse space shifted nonlocal NLS (\ref{complexspaceNLS}) in Figure 2.
\begin{center}
\begin{figure}[h!]
\centering
\begin{minipage}[t]{1\linewidth}
\centering
\includegraphics[angle=0,scale=.30]{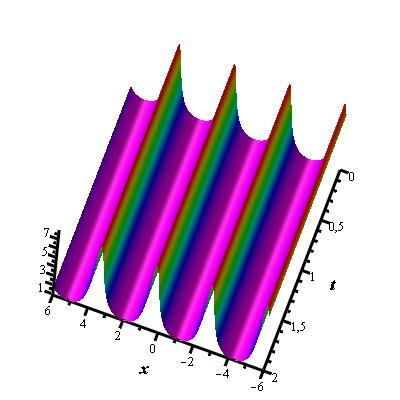}
\caption{A nonsingular periodic wave for (\ref{complexspaceNLS})
with the parameters $k=\sigma_1=\beta_1=\alpha_3=\beta_3=1, x_0=4$.}
\end{minipage}%
\end{figure}
\end{center}
\squeezeup

If we use Type 2, we obtain a different set of constraints to be satisfied by the solution parameters:
\begin{equation}\label{constraintscaseiiiType2}
k_j=\bar{k}_j,\quad Ake^{\delta_1+\bar{\delta}_1+\bar{k}_1x_0}=1, \quad Ae^{\delta_2+\bar{\delta}_2+\bar{k}_2x_0}=k,
\end{equation}
so $\omega_j=-\bar{\omega}_j$ for $j=1, 2$. Let $\omega_1=i\alpha_1$, $\omega_2=i\alpha_2$, $e^{\delta_1}=\alpha_3+i\beta_3$, $e^{\delta_2}=\alpha_4+i\beta_4$, and $a=i\alpha_5$, where $\alpha_j\in \mathbb{R}$ for $1\leq j \leq 5$, $\beta_j\in \mathbb{R}$ for $1\leq j \leq 4$. Hence one-soliton solutions of complex reverse space shifted nonlocal NLS (\ref{complexspaceNLS})
and MKdV (\ref{complexspaceMKdV}) become
\begin{equation}\label{caseiiireelsol}
|q(x,t)|^2=-\frac{e^{(k_1-k_2)x-\frac{(k_1-k_2)}{2}x_0}(k_1+k_2)^2}{2k[\cosh((k_1+k_2)x-\frac{(k_1+k_2)}{2}x_0)+\sigma_1 \cos((\alpha_1+\alpha_2)t+\omega_0)    ]},
\end{equation}
where
\begin{equation}\label{omega_0caseiii}
\omega_0=\arccos\Big(\sigma_1\frac{(\alpha_3\alpha_4-\beta_3\beta_4)}{\sqrt{(\alpha_3^2+\beta_3^2)(\alpha_4^2+\beta_4^2)}}\Big),\, \sigma_1=\pm 1,
\end{equation}
and due to  the dispersion relations, $\alpha_1=-\frac{k_1^2}{2\alpha_5}$, $\alpha_2=\frac{k_2^2}{2\alpha_5}$ for NLS, and  $\alpha_1=\frac{k_1^3}{4\alpha_5}$, $\alpha_2=\frac{k_2^3}{4\alpha_5}$ for MKdV. This solution is singular at $(x,t)=(\frac{x_0}{2},\frac{2n\pi-\omega_0}{(\alpha_1+\alpha_2)})$ if $\sigma_1=-1$ and $(x,t)=(\frac{x_0}{2},\frac{(2n+1)\pi-\omega_0}{(\alpha_1+\alpha_2)})$ if $\sigma_1=1$, $n\in \mathbb{N}$.\\

Recall the unshifted nonlocal reverse space NLS equation
\begin{equation}\label{usualnonlocalspaceNLS}\displaystyle
au_t(x,t)=\frac{1}{2}u_{xx}(x,t)-ku^2(x,t)\bar{u}(-x,t),
\end{equation}
and the unshifted nonlocal reverse space MKdV equation
\begin{equation}\label{usualnonlocalspaceMKdV}\displaystyle
au_t(x,t)=-\frac{1}{4}u_{xxx}(x,t)+\frac{3}{2}k\bar{u}(-x,t)u(x,t)u_x(x,t).
\end{equation}

When we compare the one-soliton solutions of the unshifted nonlocal reverse space NLS and MKdV equations \cite{GurPek1}, \cite{GurPek3}, \cite{GurPek2} and  the shifted nonlocal reverse space NLS (\ref{complexspaceNLS}) and MKdV (\ref{complexspaceMKdV}) equations  obtained by Hirota bilinear method via Type 1 approach, if $\alpha_1=0$ clearly we have $|q(x,t)|^2=|u(x-\frac{x_0}{2},t)|^2$.

For the solutions obtained via Type 2, it is also as if they are related with the same equality. But one must be careful about the term $\omega_0$ (\ref{omega_0caseiii}). Indeed, due to the constraints (\ref{constraintscaseiiiType2}), we have $(\alpha_3^2+\beta_3^2)(\alpha_4^2+\beta_4^2)=(k_1+k_2)^4e^{-(k_1+k_2)x_0}$. Let us consider the case when $\omega_0=0$. This happens when $\alpha_4=\tau \alpha_3$ and $\beta_4=-\tau \beta_3$ where $\tau=\pm ke^{\frac{(k_1-k_2)}{2}x_0}$. In this case the solution (\ref{caseiiireelsol}) becomes
\begin{equation}\label{omega=0soln}
|q(x,t)|^2=-\frac{e^{(k_1-k_2)x-\frac{(k_1-k_2)}{2}x_0}(k_1+k_2)^2}{2k[\cosh((k_1+k_2)x-\frac{(k_1+k_2)}{2}x_0)+\sigma_1 \cos((\alpha_1+\alpha_2)t)]},
\end{equation}
where $\sigma_1=\pm 1$. Surely, we have the relation $|q(x,t)|^2=|u(x-\frac{x_0}{2},t)|^2$ between the above solution (\ref{omega=0soln}) of shifted nonlocal reverse space NLS (\ref{complexspaceNLS}) and MKdV (\ref{complexspaceMKdV}) equations, and  the one-soliton solutions of unshifted nonlocal reverse space NLS and MKdV equations obtained by Type 2. This solution is singular at $(x,t)=(\frac{x_0}{2},\frac{2n\pi}{(\alpha_1+\alpha_2)})$ if $\sigma_1=-1$ and $(x,t)=(\frac{x_0}{2},\frac{(2n+1)\pi}{(\alpha_1+\alpha_2)})$ if $\sigma_1=1$, $n\in \mathbb{N}$.\\

In \cite{AbMu4}, Ablowitz and Musslimani gave one-soliton solution to the complex reverse space shifted nonlocal NLS (\ref{complexspaceNLS}) with $ a=\frac{i}{2}$ as
\begin{equation}\displaystyle
q(x,t)=-\frac{2(\eta_1+\bar{\eta}_1)e^{2\eta_1x-4i\eta_1^2t-i\theta_1-\eta_1x_0}}{1-e^{2(\bar{\eta}_1+\eta_1)x+4i(\bar{\eta}_1^2-\eta_1^2)t-
i(\theta_1+\bar{\theta}_1)-(\eta_1+\bar{\eta}_1)x_0}}.
\end{equation}
Here $\eta_1>0$, $\bar{\eta}_1>0$, $\theta_1$, and $\bar{\theta}_1$ are four free real parameters. They explored that the above solution that they obtained by using inverse scattering transform via Riemann-Hilbert formulation, is actually related to the solution, say $u(x,t)$, of the focusing complex reverse space nonlocal NLS equation by the relation $q(x,t)=u(x-\frac{x_0}{2},t)$.\\

\noindent \textbf{iv)}\, $r(x,t)=k\bar{q}(x,-t+t_0)$, $k, t_0 \in \mathbb{R}$.\\

When we use Type 1 approach here we get
\begin{equation}
k_2=\bar{k}_1,\quad e^{\delta_2}=ke^{\bar{\delta}_1+\bar{\omega}_1t_0},
\end{equation}
giving $\omega_2=-\bar{\omega}_1$. Hence we obtain one-soliton solutions of  reverse time shifted nonlocal NLS (\ref{complextimeNLS}) and
MKdV (\ref{complextimeMKdV}) equations
as
\begin{equation}\label{caseivoneType1}\displaystyle
q(x,t)=\frac{e^{k_1x+\omega_1t+\delta_1}}{1-\frac{k}{(k_1+\bar{k}_1)^2}e^{(k_1+\bar{k}_1)x+(\omega_1-\bar{\omega}_1)t+\delta_1+\bar{\delta}_1+\bar{\omega}_1t_0}},
\end{equation}
where $\omega_1=\frac{k_1^2}{2a}$, $a=\bar{a}$ for NLS and $\omega_1=-\frac{k_1^3}{4a}$, $a=-\bar{a}$ for MKdV equations. Let $k_1=\alpha_1+i\beta_1$, $\omega_1=\alpha_2+i\beta_2$ and $e^{\delta_1}=\alpha_3+i\beta_3$ for $\alpha_j, \beta_j \in \mathbb{R}$, $j=1,2,3$. For NLS the constant $a$ is real, but for MKdV it is pure imaginary, say $a=i\alpha_4$, $\alpha_4\in \mathbb{R}$. Therefore $\alpha_2=\frac{(\alpha_1^2-\beta_1^2)}{2a}, \beta_2=\frac{\alpha_1
\beta_1}{a}$ for NLS and $\alpha_2=\frac{(\beta_1^3-3\alpha_1^2\beta_1)}{4\alpha_4}$, $\beta_2=\frac{(\alpha_1^3-3\alpha_1\beta_1^2)}{4\alpha_4}$ for MKdV.
Then if $\beta_1\neq 0$ the solution (\ref{caseivoneType1}) becomes
\begin{equation}\label{caseivoneType1}
|q(x,t)|^2=\frac{2e^{2\alpha_2t-\alpha_2t_0}}{k\sigma_1[\cosh(2\alpha_1x+\alpha_2t_0+\delta)-\sigma_1\cos(2\beta_2t-\beta_2t_0)]},
\end{equation}
where $\delta=\ln\Big(\frac{k\sigma_1(\alpha_3^2+\beta_3^2)}{4\alpha_1^2}\Big)$. Here if $k<0$, $\sigma_1=-1$, and $k>0$, $\sigma_1=1$. If $k<0$, the above solution is singular at $(x,t)=(-\frac{(\alpha_2t_0+\delta)}{2\alpha_1}, \frac{(2n+1)\pi+\beta_2t_0}{2\beta_2})$, $n\in \mathbb{N}$. If $k>0$, then the solution is singular at $(x,t)=(-\frac{(\alpha_2t_0+\delta)}{2\alpha_1}, \frac{2n\pi+\beta_2t_0}{2\beta_2})$, $n\in \mathbb{N}$.

If $\beta_1=0$ then $\beta_2=0$ for NLS. Then the solution (\ref{caseivoneType1}) is nonsingular for $k<0$ i.e. $\sigma=-1$. If $u(x,t)$ is the solution of reverse time unshifted nonlocal NLS, we have $q(x,t)\neq u(x,t+\kappa(t_0))$ for any nonzero function $\kappa$. However, for MKdV $\beta_1=0$ yields $\alpha_2=0$, and the solution (\ref{caseivoneType1}) becomes still singular but we have the relation $|q(x,t)|^2=|u(x,t-\frac{t_0}{2})|^2$ where $u(x,t)$ is the solution of
reverse time unshifted nonlocal MKdV equation obtained by Type 1.

If $\beta_1=\pm \alpha_1$ for NLS, $\beta_1=\pm \sqrt{3}\alpha_1$ for MKdV, we have $\alpha_2=0$. Under these conditions the solution (\ref{caseivoneType1}) is again singular but we have $|q(x,t)|^2=|u(x,t-\frac{t_0}{2})|^2$ where $u(x,t)$ is the solution of
unshifted nonlocal NLS and MKdV equations derived by Type 1.\\

Using Type 2 gives a different solution where the solution parameters are satisfying;
\begin{equation}
k_j=-\bar{k}_j,\quad Ake^{\delta_1+\bar{\delta}_1+\bar{\omega}_1t_0}=1, \quad Ae^{\delta_2+\bar{\delta}_2+\bar{\omega}_2t_0}=k,
\end{equation}
so $\omega_j=\bar{\omega}_j$ for $j=1, 2$. Let $k_1=i\alpha_1$, $k_2=i\alpha_2$, $e^{\delta_1}=\alpha_3+i\beta_3$, and $e^{\delta_2}=\alpha_4+i\beta_4$,
where $\alpha_j \in \mathbb{R}$ for $1\leq j \leq 4$ and $\beta_3, \beta_4 \in \mathbb{R}$. Here $a\in \mathbb{R}$ for NLS and $a=i\alpha_5$, $\alpha_5\in \mathbb{R}$ for MKdV. Hence we obtain one-soliton solutions of the reverse time shifted nonlocal NLS (\ref{complextimeNLS}) and
MKdV (\ref{complextimeMKdV}) equations as
\begin{equation}\label{caseivreelsol}
|q(x,t)|^2=\frac{e^{(\omega_1-\omega_2)t-\frac{(\omega_1-\omega_2)}{2}t_0}(\alpha_1^2+\alpha_2^2)}{2k[\cosh((\omega_1+\omega_2)t-\frac{(\omega_1+\omega_2)}{2}t_0  )+\sigma_1\cos((\alpha_1+\alpha_2)x+\omega_0)]},
\end{equation}
where $\sigma_1=\pm 1$, $\omega_0=\arccos\Big(\frac{(\alpha_3\alpha_4-\beta_3\beta_4)e^{\frac{(\omega_1+\omega_2)t_0}{2}}}{(\alpha_1+\alpha_2)^2}     \Big)$, and due to the dispersion relations $\omega_1=-\frac{\alpha_1^2}{2a}$, $\omega_2=\frac{\alpha_2^2}{2a}$ for NLS, and $\omega_1=\frac{\alpha_1^3}{4\alpha_5}$, $\omega_2=\frac{\alpha_2^3}{4\alpha_5}$ for MKdV. The solution (\ref{caseivreelsol}) is singular at
$(x,t)=(\frac{(2n\pi-\omega_0)}{(\alpha_1+\alpha_2)}, \frac{t_0}{2})$ if $\sigma_1=-1$ and at $(x,t)=(\frac{((2n+1)\pi-\omega_0)}{(\alpha_1+\alpha_2)}, \frac{t_0}{2})$ if $\sigma_1=1$, $n \in \mathbb{N}$.

Consider the case when $\omega_0=0$ which happens when $\alpha_4=\tau \alpha_3$ and $\beta_4=-\tau \beta_3$ where $\tau=\pm ke^{\frac{(\omega_1-\omega_2)}{2}t_0}$. Under this condition the solution (\ref{caseivreelsol}) of reverse time shifted nonlocal NLS (\ref{complextimeNLS}) and MKdV (\ref{complextimeMKdV}) equations is still singular but we have the relation $|q(x,t)|^2=|u(x,t-\frac{t_0}{2})|^2$ where $u(x,t)$ is one-soliton solution of reverse time unshifted nonlocal NLS and MKdV equations obtained by Type 2.\\

\noindent \textbf{v)}\, $r(x,t)=k\bar{q}(-x+x_0,-t+t_0)$, $k, x_0, t_0 \in \mathbb{R}$.\\

Using Type 1 yields
\begin{equation}
k_2=-\bar{k}_1,\quad e^{\delta_2}=ke^{\bar{\delta}_1+\bar{k}_1x_0+\bar{\omega}_1t_0},
\end{equation}
giving $\omega_2=-\bar{\omega}_1$. Under these constraints we obtain one-soliton solutions of  reverse space-time shifted nonlocal NLS  (\ref{complexspacetimeNLS}) and MKdV (\ref{complexspacetimeMKdV}) equations as
\begin{equation}\label{casevoneType1}\displaystyle
q(x,t)=\frac{e^{k_1x+\omega_1t+\delta_1}}{1-\frac{k}{(k_1-\bar{k}_1)^2}e^{(k_1-\bar{k}_1)x+(\omega_1-\bar{\omega}_1)t+\delta_1
+\bar{\delta}_1+\bar{k}_1x_0+\bar{\omega}_1t_0}},
\end{equation}
where $\omega_1=\frac{k_1^2}{2a}$ for NLS and $\omega_1=-\frac{k_1^3}{4a}$ for MKdV equations. Here the constant $a$ is a real number. Let $k_1=\alpha_1+i\beta_1$, $\omega_1=\alpha_2+i\beta_2$ and $e^{\delta_1}=\alpha_3+i\beta_3$ for $\alpha_j, \beta_j \in \mathbb{R}$, $j=1,2,3$. In this case $\alpha_2=\frac{(\alpha_1^2-\beta_1^2)}{2a}, \beta_2=\frac{\alpha_1
\beta_1}{a}$ for NLS and $\alpha_2=\frac{(3\alpha_1\beta_1^2-\alpha_1^3)}{4a}$, $\beta_2=\frac{(\beta_1^3-3\alpha_1^2\beta_1)}{4a}$ for MKdV. Hence if $\alpha_1\neq 0$ from the solution (\ref{casevoneType1}) we have
\begin{equation}\label{casevreeloneType1}
|q(x,t)|^2=\frac{2\beta_1^2e^{2\alpha_1x+2\alpha_2t-\alpha_1x_0-\alpha_2t_0}}{k\sigma_1[\cosh(\alpha_1x_0+\alpha_2t_0+\delta)+\sigma_1\cos(2\beta_1x
+2\beta_2t-\beta_1x_0-\beta_2t_0)]},
\end{equation}
where $\delta=\ln\Big(\frac{k\sigma_1(\alpha_3^2+\beta_3^2)}{4\beta_1^2}\Big)$. Here if $k<0$, $\sigma_1=-1$, and $k>0$, $\sigma_1=1$. If $\alpha_1x_0+\alpha_2t_0+\delta\neq 0$  then the solution is nonsingular.

If $\alpha_1=0$ then $\alpha_2=0$ for MKdV. In this case the solution (\ref{casevreeloneType1}) is nonsingular if $\delta>0$ and
we have the relation $|q(x,t)|^2=|u(x-\frac{x_0}{2},t-\frac{t_0}{2})|^2$ where $u(x,t)$ is one-soliton solution of reverse
space-time unshifted nonlocal MKdV equation obtained by Type 1.

If we use Type 2, we obtain a different solution where the solution parameters are satisfying;
\begin{equation}
k_j=\bar{k}_j,\quad Ake^{\delta_1+\bar{\delta}_1+\bar{k}_1x_0+\bar{\omega}_1t_0}=1, \quad Ae^{\delta_2+\bar{\delta}_2+\bar{k}_2x_0+\bar{\omega}_2t_0}=k,
\end{equation}
so $\omega_j=\bar{\omega}_j$ for $j=1, 2$. Let $e^{\delta_1}=\alpha_1+i\beta_1$ and $e^{\delta_2}=\alpha_2+i\beta_2$, where $\alpha_j, \beta_j \in \mathbb{R}$ for $j=1, 2$. Then we get one-soliton solutions of  reverse space-time shifted nonlocal NLS  (\ref{complexspacetimeNLS}) and MKdV (\ref{complexspacetimeMKdV}) equations as
\begin{equation}\label{casevreelsol}\displaystyle
|q(x,t)|^2=\frac{e^{2k_1x+2\omega_1t}(\alpha_1^2+\beta_1^2)}{1-2\gamma_1e^{\psi}+\gamma_2e^{2\psi}},
\end{equation}
where
\begin{equation}\displaystyle
\psi=(k_1+k_2)x+(\omega_1+\omega_2)t,\, \gamma_1=\frac{(\alpha_1\alpha_2-\beta_1\beta_2)}{(k_1+k_2)^2},\, \gamma_2=e^{(k_1+k_2)x_0+(\omega_1+\omega_2)t_0}.
\end{equation}
The solution (\ref{casevreelsol}) becomes singular when the denominator $f(\psi)=1-2\gamma_1e^{\psi}+\gamma_2e^{2\psi}$ vanishes.
This function becomes zero when $e^{\psi}=\frac{\gamma_1\pm \sqrt{\gamma_1^2-\gamma_2}}{\gamma_2}$. Hence if $\gamma_1^2<\gamma_2$,
the solution (\ref{casevreelsol}) is nonsingular. We can rewrite the solution (\ref{casevreelsol}) as
\begin{equation}\label{casevoneType2}\displaystyle
|q(x,t)|^2=\frac{-e^{(k_1-k_2)x+(\omega_1-\omega_2)t-\frac{(k_1-k_2)}{2}x_0-\frac{(\omega_1-\omega_2)}{2}t_0}(k_1+k_2)^2}
{2k[\cosh((k_1+k_2)x+(\omega_1+\omega_2)t-\frac{(k_1+k_2)}{2}x_0-\frac{(\omega_1+\omega_2)}{2}t_0)-\rho]},
\end{equation}
where
\begin{equation}\displaystyle
\rho=\frac{(\alpha_1\alpha_2-\beta_1\beta_2)}{(k_1+k_2)^2}e^{\frac{(k_1+k_2)}{2}x_0+\frac{(\omega_1+\omega_2)}{2}t_0}.
\end{equation}
Consider the case when $\rho=0$. This happens when $\alpha_2=\tau\beta_1$ and $\beta_2=\tau \alpha_1$ where $\tau=\pm k e^{ \frac{(k_1-k_2)}{2}x_0+\frac{(\omega_1-\omega_2)}{2}t_0}$. In this case the solution (\ref{casevoneType2}) becomes nonsingular for any $(x,t)\in \mathbb{R}^2$
and the relation $|q(x,t)|^2=|u(x-\frac{x_0}{2},t-\frac{t_0}{2})|^2$ holds where $u(x,t)$ is one-soliton solution of reverse
space-time unshifted nonlocal NLS and MKdV equations obtained by Type 2.

\subsection{Two-soliton solutions}

By using the Hirota bilinear method we obtained two-soliton solutions of the coupled NLS system (\ref{NLS1}) and (\ref{NLS2}) \cite{GurPek1}, \cite{GurPek3}
and MKdV system (\ref{MKdV1}) and (\ref{MKdV2}) in  \cite{GurPek3}, and \cite{GurPek2}. Similar to one-soliton solutions, we can also obtain two-soliton solutions of the shifted nonlocal NLS and MKdV equations by using Type 1 and Type 2 approaches with the reduction formulas.\\

\noindent \textbf{i)}\, $r(x,t)=kq(x,-t+t_0)$, $k, t_0 \in \mathbb{R}$.\\

Using Type 1 with this reduction formula for the coupled NLS system we get
the following constraints:
\begin{equation}
l_j=k_j,\quad e^{\alpha_j}=ke^{\delta_j+\omega_jt_0}
\end{equation}
giving $\omega_j=-m_j$ for $j=1, 2$. Here the dispersion relation (\ref{NLSdispersiontwo}) holds. Hence we get two-soliton solution of real reverse time shifted nonlocal NLS equation (\ref{realtimeNLS}) as
(\ref{twosolq(x,t)}) and (\ref{twosolr(x,t)}) satisfying the above conditions.\\

\noindent \textbf{ii)}\, $r(x,t)=kq(-x+x_0,-t+t_0)$, $k, x_0, t_0 \in \mathbb{R}$.\\

When we use Type 1 in this case for both NLS and MKdV equations we get trivial solution $q(x,t)=0$. Hence we use Type 2 and get the following constraints:
\begin{equation}\displaystyle
e^{\delta_j}=\tau_r i\frac{(k_j+l_1)(k_j+l_2)}{\sqrt{k}(k_1-k_2)e^{\frac{(k_jx_0+\omega_jt_0)}{2}}},\quad e^{\alpha_j}=\rho_r i\sqrt{k}\frac{(k_1+l_j)(k_2+l_j)}{(l_1-l_2)e^{\frac{(l_jx_0+m_jt_0)}{2}}}, \tau_r=\pm 1, \rho_r=\pm 1, r=1, 2,
\end{equation}
 for $j=1, 2$. Therefore two-soliton solutions of real reverse space-time shifted nonlocal NLS (\ref{realspacetimeNLS}) and MKdV (\ref{realspacetimeMKdV}) equations are given by (\ref{twosolq(x,t)}) and (\ref{twosolr(x,t)}) satisfying the above constraints and the dispersion relations (\ref{NLSdispersiontwo}) and (\ref{MKdVdispersiontwo}), respectively.\\

\noindent \textbf{iii)}\, $r(x,t)=k\bar{q}(-x+x_0,t)$, $k, x_0 \in \mathbb{R}$.\\

Here Type 1 yields
\begin{equation}\label{iiitwoconst}
l_j=-\bar{k}_j,\quad e^{\alpha_j}=ke^{\bar{\delta}_j+\bar{k}_jx_0},
\end{equation}
giving $m_j=\bar{\omega}_j$ for $j=1, 2$. Hence we obtained two-soliton solutions of complex reverse space shifted nonlocal NLS (\ref{complexspaceNLS})
and MKdV (\ref{complexspaceMKdV}) as in (\ref{twosolq(x,t)}) and (\ref{twosolr(x,t)}) where the parameters of the solutions are satisfying (\ref{iiitwoconst}) with the dispersion relations (\ref{NLSdispersiontwo}) for NLS and (\ref{MKdVdispersiontwo}) for MKdV equations. Here $a$ is a pure imaginary number. Let us write the two-soliton solution obtained here explicitly;
\begin{equation}
q(x,t)=\frac{\zeta(x,t)}{\phi(x,t)},
\end{equation}
where
\begin{align}
&\zeta(x,t)=e^{k_1x+\omega_1t+\delta_1}+e^{k_2x+\omega_2t+\delta_2}+A_1ke^{(k_1+k_2-\bar{k}_1)x+(\omega_1+\omega_2+\bar{\omega}_1)t}e^{\delta_1+\delta_2+\bar{\delta}_1+\bar{k}_1x_0}\nonumber\\
&+A_2ke^{(k_1+k_2-\bar{k}_2)x+(\omega_1+\omega_2+\bar{\omega}_2)t}e^{\delta_1+\delta_2+\bar{\delta}_2+\bar{k}_2x_0}\\
&\phi(x,t)=1+ke^{(k_1-\bar{k}_1)x+(\omega_1+\bar{\omega}_1)t}e^{\delta_1+\bar{\delta}_1+\bar{k}_1x_0+\alpha_{11}}+
ke^{(k_1-\bar{k}_2)x+(\omega_1+\bar{\omega}_2)t}e^{\delta_1+\bar{\delta}_2+\bar{k}_2x_0+\alpha_{12}}\nonumber\\
&+ke^{(k_2-\bar{k}_1)x+(\omega_2+\bar{\omega}_1)t}e^{\delta_2+\bar{\delta}_1+\bar{k}_1x_0+\alpha_{21}}
+ke^{(k_2-\bar{k}_2)x+(\omega_2+\bar{\omega}_2)t}e^{\delta_2+\bar{\delta}_2+\bar{k}_2x_0+\alpha_{22}}\nonumber\\
&+Mk^2e^{(k_1+k_2-\bar{k}_1-\bar{k}_2)x+(\omega_1+\omega_2+\bar{\omega}_1+\bar{\omega}_2)t}e^{\delta_1+\delta_2+\bar{\delta}_1+\bar{\delta}_2
+(\bar{k}_1+\bar{k}_2)x_0},
\end{align}
with
\begin{equation}\displaystyle
e^{\alpha_{ij}}=-\frac{1}{(k_i-\bar{k}_j)^2},\, 1\leq i,j\leq 2,\, \, A_i=-\frac{(k_1-k_2)^2}{(k_1-\bar{k}_i)^2(k_2-\bar{k}_i)^2},\, \, i=1, 2,
\end{equation}
and
\begin{equation}\displaystyle
M=\frac{(k_1-k_2)^2(\bar{k}_1-\bar{k}_2)^2}{(k_1-\bar{k}_1)^2(k_1-\bar{k}_2)^2(k_2-\bar{k}_1)^2(k_2-\bar{k}_2)^2}.
\end{equation}
Here $ \omega_j=\frac{k_j^2}{2a}$, $j=1, 2$ for NLS and $ \omega_j=-\frac{k_j^3}{4a}$, $j=1, 2$ for MKdV. \\

\noindent \textbf{Example 3.} Choose the parameters as $k_1=\frac{i}{4}, k_2=\frac{i}{2}, a=2i, k=-1, \delta_1=\delta_2=0$. For $x_0=-4$ we give the graph of the two-soliton solution of complex reverse space shifted nonlocal NLS (\ref{complexspaceNLS}) in Figure 3.
\begin{center}
\begin{figure}[h!]
\centering
\begin{minipage}[t]{1\linewidth}
\centering
\includegraphics[angle=0,scale=.30]{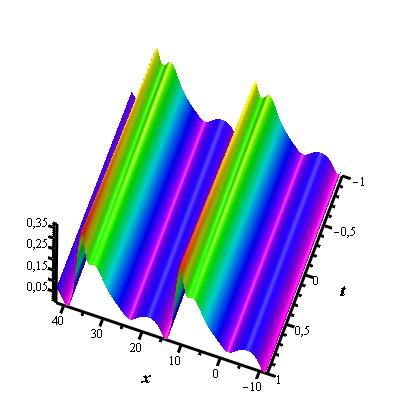}
\caption{Periodic wave solution for (\ref{complexspaceNLS})
with the parameters $k_1=\frac{i}{4}, k_2=\frac{i}{2}, a=2i, k=-1, \delta_1=\delta_2=0, x_0=-4$.}
\end{minipage}%
\end{figure}
\end{center}
\squeezeup

\noindent \textbf{iv)}\, $r(x,t)=k\bar{q}(x,-t+t_0)$, $k, t_0 \in \mathbb{R}$.\\

If we use Type 1 approach we get
\begin{equation}
l_j=k_j,\quad e^{\alpha_j}=ke^{\bar{\delta}_j+\bar{\omega}_jt_0},
\end{equation}
giving $m_j=-\bar{\omega}_j$ for $j=1, 2$. Hence two-soliton solutions of  reverse time shifted nonlocal NLS (\ref{complextimeNLS}) and
MKdV (\ref{complextimeMKdV}) equations are given by (\ref{twosolq(x,t)}) and (\ref{twosolr(x,t)}) where the parameters of the solutions are satisfying the
above constraints with the dispersion relations (\ref{NLSdispersiontwo}) and $a=\bar{a}$  for NLS, and (\ref{MKdVdispersiontwo}), $a=-\bar{a}$ for MKdV equations.\\

\noindent \textbf{v)}\, $r(x,t)=k\bar{q}(-x+x_0,-t+t_0)$, $k, x_0, t_0 \in \mathbb{R}$.\\

In this case Type 1 yields
\begin{equation}
l_j=-k_j,\quad e^{\alpha_j}=ke^{\bar{\delta}_j+\bar{k}_jx_0+\bar{\omega}_jt_0},
\end{equation}
giving $m_j=-\bar{\omega}_j$ for $j=1, 2$. So two-soliton solutions of  reverse space-time shifted nonlocal NLS  (\ref{complexspacetimeNLS}) and MKdV (\ref{complexspacetimeMKdV}) equations are given by (\ref{twosolq(x,t)}) and (\ref{twosolr(x,t)}) where the parameters are satisfying the
above constraints with the dispersion relations (\ref{NLSdispersiontwo}) for NLS, and (\ref{MKdVdispersiontwo}) for MKdV equations. Here the constant $a$ is a real number.\\

\section{Conclusion}
In this work we presented all consistent shifted nonlocal reductions of NLS and MKdV systems. By using
one- and two-soliton solutions of the coupled NLS and MKdV system of equations and the shifted nonlocal reduction formulas we obtained one- and two-soliton solutions of the shifted nonlocal NLS and MKdV equations. We compared our one-soliton solutions with those obtained previously for unshifted nonlocal NLS and MKdV equations.

\section{Acknowledgment}
  This work is partially supported by the Scientific
and Technological Research Council of Turkey (T\"{U}B\.{I}TAK).


\begin{thebibliography}{99}

\bibitem{AbMu1} M.J. Ablowitz and Z.H. Musslimani, Integrable nonlocal nonlinear Schr\"{o}dinger equation, Phys. Rev. Lett. 110 (2013) 064105.

\bibitem{AbMu2} M.J. Ablowitz and Z.H. Musslimani, Inverse scattering transform for the integrable nonlocal nonlinear Schr\"{o}dinger equation, Nonlinearity 29 (2016) 915--946.

\bibitem{AbMu3} M.J. Ablowitz and Z.H. Musslimani, Integrable nonlocal nonlinear equations, Stud. App. Math. 139 (1) (2016) 7--59.


\bibitem{AbMu4} M.J. Ablowitz and Z.H. Musslimani, Integrable space-time shifted nonlocal nonlinear equations, Phys. Lett. A \textbf{409}, 127516, 2021.

\bibitem{chen} K. Chen, X. Deng, S. Lou, and D. Zhang, Solutions of local and nonlocal equations reduced from the AKNS hierarchy, Stud. Appl. Math. 141 (1) (2018) 113--141.

\bibitem{FLAH} B.F. Feng, X.D. Luo, M.J. Ablowitz, and Z.H. Musslimani, General soliton solution to a nonlocal nonlinear Schr\"{o}dinger
equation with zero and nonzero boundary conditions, Nonlinearity 31 (12) (2018) 5385--5409.


\bibitem{gerd} V.S. Gerdjikov and A. Saxena, Complete integrability of nonlocal nonlinear Schr\"{o}dinger equation, J. Math. Phys. 58 (1) (2017) 013502.



\bibitem{GurPek1}  M. G\"{u}rses and A. Pekcan, Nonlocal nonlinear Schr\"{o}dinger equations and their soliton solutions, J. Math. Phys. 59 (2018) 051501.

\bibitem{GurPek3} M. G\"{u}rses and A. Pekcan, Integrable nonlocal reductions, Symmetries, Differential Equations and Applications SDEA-III, Istanbul, Turkey, August 2017, in: V.G. Kac, P.J. Olver, P. Winternitz, T. Ozer (Eds), Springer Proceedings in Mathematics and Statistics, No: 266, 2018, pp: 27-52.


\bibitem{huang} X. Huang, L. Ling, Soliton solutions for the nonlocal nonlinear Schr\"{o}dinger equation, Eur. Phys. J. Plus 131 (2016) 148.


\bibitem{li}  M. Li and T. Xu, Dark and antidark soliton interactions in the nonlocal nonlinear Schr\"{o}dinger equation
with the self-induced parity-time-symmetric potential, Phys. Rev. E 91 (2015) 033202.


\bibitem{aflm1} M.J. Ablowitz, B.F. Feng, X.D. Luo, and Z.H. Musslimani, Inverse scattering transform for the nonlocal reverse space-time
nonlinear Schr\"{o}dinger equation, Theor. Math. Phys. 196 (3) (2018) 1241--1267.


\bibitem{Wen} X.Y. Wen, Z. Yan, and Y. Yang, Dynamics of higher-order rational solitons for the nonlocal nonlinear Schr\"{o}dinger
equation with the self-induced parity-time-symmetric potential, Chaos 26 (2015) 063123.

\bibitem{Sax} A. Khare and A. Saxena, Periodic and hyperbolic soliton solutions of a number of nonlocal nonlinear
equations, J. Math. Phys. 56 (2015) 032104.

\bibitem{XLLLZ} T. Xu, S. Lan, M. Li, L.-L. Li, and G.-W. Zhang, Mixed soliton solutions of the defocusing nonlocal nonlinear Schr\"{o}dinger equation, Phys. D
\textbf{390}, 47--61, 2019.

\bibitem{XCLM} T. Xu, Y. Chen, M. Li, and D.-X. Meng, General stationary solutions of the nonlocal nonlinear Schr\"{o}dinger equation and their relevance
to the $\mathcal{PT}$-symmetric system, Chaos \textbf{29}, 123124, 2019.

\bibitem{Ma1} W.-X. Ma, Inverse scattering for nonlocal reverse-time nonlinear Schr\"{o}dinger equations, Appl. Math. Lett. \textbf{102}, 106161, 2020.


\bibitem{jianke} J. Yang, General N-solitons and their dynamics in several nonlocal nonlinear Schr\"{o}dinger equations, Phys. Lett A 383 (4) (2019) 328--337.
\bibitem{fok} A.S. Fokas, Integrable multidimensional versions of the nonlocal Schr\"{o}dinger equation, Nonlinearity 29 (2016) 319.
\bibitem{sin} D. Sinha and P.K. Ghosh, Integrable nonlocal vector nonlinear Schr\"{o}dinger equation with self-induced parity-time symmetric potential, Phys. Lett. A 381 (2017) 124.

      \bibitem{zyan} Z. Yan, Integrable PT-symmetric local and nonlocal vector nonlinear  Schr\"{o}dinger equations: A unified two parameter model, Appl. Math. Lett. 47 (2015) 61






\bibitem{GurPek2}  M. G{\" u}rses and A. Pekcan, Nonlocal nonlinear modified KdV equations and their soliton solutions, Commun. Nonlinear Sci. Numer.
Simul. 2019; 67: 427.

\bibitem{JZ1} J.L. Ji and Z.N. Zhu, On a nonlocal modified Korteweg-de Vries equation: Integrability, Darboux transformation and soliton solutions,
Commun. Non. Sci. Numer. Simul. 42 (2017) 699.

\bibitem{JZ2} J.L. Ji and Z.N. Zhu, Soliton solutions of an integrable nonlocal modified Korteweg-de Vries equation through
inverse scattering transform, J. Math. Anal. Appl. 453 (2017) 973.

\bibitem{Shi} X. Shi, P. Lv, and C. Qi, Explicit solutions to a nonlocal 2-component complex modified Korteweg-de Vries equation, Appl. Math. Lett.
\textbf{100}, 106043, 2020.

\bibitem{Luo} X.-D. Luo, Inverse scattering transform for the complex reverse space-time nonlocal modified Korteweg-de Vries equation with nonzero boundary conditions and constant phase shift, Chaos \textbf{29}, 073118, 2019.

\bibitem{Yan} G. Zhang, Z. Yan, Inverse scattering transforms and soliton solutions of focusing and defocusing nonlocal mKdV equations with nonzero boundary conditions, Phys. D \textbf{402}, 132170, 2020.

\bibitem{ma}   L.Y. Ma, S.F. Shen, and Z.N. Zhu, Soliton solution and gauge equivalence for an integrable nonlocal complex modified Korteweg-de Vries
equation, J. Math. Phys. 2017; 58: 103501.

\bibitem{aflm} M.J. Ablowitz, B.F. Feng, X.D. Luo, and Z.H. Musslimani, Reverse space-time nonlocal sine-Gordon/sinh-Gordon equations with nonzero boundary conditions, Stud. Appl. Math. 141 (3) (2018) 267--307.







\bibitem{RZFH} J. Rao, Y. Zhang, A.S. Fokas, and J. He, Rogue waves of the nonlocal Davey-Stewartson I equation, Nonlinearity 31 (2018) 4090--4107.

\bibitem{XLHCY} T. Xu, M. Li, Y. Huang, Y. Chen, and C. Yu, Nonsingular localized wave solutions for the nonlocal Davey-Stewartson I equation
with zero background, Modern Phys. Lett. B 31 (35) (2017) 1750338.

\bibitem{ZXZhou} Z.-X. Zhou, Darboux transformations global explicit solutions for nonlocal Davey-Stewartson I equation, Stud. Appl. Math. 141 (2) (2018) 186--204.


\bibitem{ZL} Y. Zhang and Y. Liu, Breather and lump solutions for nonlocal Davey-Stewartson II equation, Nonlinear Dyn. 96 (2019) 107--113.

\bibitem{Rao} J. Rao, Y. Cheng, and J. He, Rational and semirational solutions of the nonlocal Davey-Stewartson equations, Stud. Appl. Math. \textbf{139} (4), 2017.



\bibitem{GursesFK} M. G\"{u}rses, Nonlocal Fordy-Kulish equations on symmetric spaces, Phys. Lett. A 381 (2017) 1791.


\bibitem{gerd2} V.S. Gerdjikov, G.G. Grahovski, and R.I. Ivanov, On the N-wave equations with PT symmetry, Theor. and Math. Phys. 188 (3) (2016) 1305.





 \bibitem{gerd1}  V.S. Gerdjikov, G.G. Grahovski, and R.I. Ivanov, On integrable wave interactions and Lax pairs on symmetric spaces, Wave Motion 71 (2017) 53.

     \bibitem{gerd3}   V.S. Gerdjikov, On nonlocal models of Kulish-Sklyanin type and generalized Fourier transforms, Stud. Comp. Int. 681 (2017) 37.


\bibitem{GurPek4} M. G\"{u}rses and A. Pekcan, $(2+1)$-dimensional local and nonlocal reductions of the
negative AKNS system: Soliton solutions, Commun. Nonlinear Sci. Numer. Simul. 71 (2019) 161--173.

\bibitem{Pek} A. Pekcan,  Nonlocal coupled HI-MKdV systems, Commun. Nonlinear Sci. Numer. Simul. 72 (2019) 493--515.

\bibitem{GurPek5} M. G\"{u}rses, A. Pekcan, (2+1)-dimensional AKNS(-N) systems II, Commun. Nonlinear Sci. Numer. Simul. \textbf{97}, 105736, 2021.

\bibitem{GurPek6} M. G\"{u}rses, A. Pekcan, Nonlocal KdV equations, Phys. Lett. A \textbf{384} (35), 126894, 2020.

\bibitem{hydro} M. G{\" u}rses, A. Pekcan, and K. Zheltukhin, Nonlocal hydrodynamic type of equations, Commun. Nonlinear Sci. Numer. Simul. 85 (2020) 105242.


























\end{thebibliography}
\end{document}